\documentclass[11pt,preprint]{aastex}
\usepackage{epsfig}
\begin{document}

\title{Probe into the interior of compact stars}
\author{Nana Pan$^{1}$, Miao Kang$^{2}$\\
1 - {\it The institute of astrophysics,Huazhong normal
university},  Wuhan 430079, Hubei, P.R.China.\\
 2 - {\it  The college of
physics and electron,Henan university}, Kaifeng, 475004, Henan,
P.R.China.}

%\classification{97.60Jd,12.38.Mh,97.60.Gb}\\
 %\keywords{dense
%matter --- equation of state --- stars: evolution--- stars:
%neutron --- stars: rotation}

\begin{abstract}
The interior of neutron stars contains nuclear matter at very high
density for numerous subatomic particles compete with each other.
Therefore, confirming the components and properties there is our
significant task. Here we summarize  the possible methods especial
the way of r-mode instability to probe into the neutron star and
show our some results. The KHz pulsar in XTE J1739-285 may give a
significant implication.
\end{abstract}

\maketitle

Since the idea of neutron star was proposed and then the concept
that pulsars are rotating neutron star was accepted, the
components and properties of interiors in neutron stars have
attracted much attention. Neutron stars can actually be
constituted of a large variety of particles other than just
neutrons and protons, therefore, the equations of state (EOSs) are
still an indeterminacy in investigation just due to the
uncertainties of nuclear physics theories.

Many investigators spontaneously expect to probe the matter under
these conditions through astrophysical observations. Up to now,
constraints of inferred masses and radii for pulsars with
sufficient precision on EOSs cannot uniquely examine and
distinguish the compositions inside neutron stars(\cite{Pan:2007})
. Although \"{O}zel had ever claimed
 that the existence of condensates and unconfined quarks may be ruled out by the mass M and radius R of
 the neutron star EXO 0748-676(\cite{Ozel:2006}), Alford et al. immediately disproved it(\cite{Alford:2007}).
 Along with the abundance of data about the spin frequency for pulsars,
researchers begin to pay much attention to the constraints put by
limit rotation. For a uniform compact star with  M and R, its
rotation frequency has an absolute mass-shedding limit
\begin{equation}
\nu_{K}=1.042\times1000\frac{(M/M_{\bigodot})^{1/2}}{(R/10km)^{3/2}}
Hz, \label{KC}
\end{equation}
 above which the star is unstable.
It is foreign to the properties of the matter inside. After giving
the frequency of observed pulsar and assuming it to be the
Keplerian limit, we can easily obtain a M-R relation to constrain
the EOSs and get an upper limit for the R of the compact stars
(CSs). But unfortunately, past works have showed that CSs
containing exotic particles are indistinguishable from traditional
neutron stars (NSs) only by this method. Actually, gravitational
waves instabilities may occur just before the rotation of CS could
reach the Keplerian limit. There exists various oscillation modes
in the CSs, while Andersson, Friedman and Morsink discovered that
only the r-mode is the most important. It could cause the
strongest instabilities of gravitational waves and plays a key
role in determining whether rotating CSs are stable or not. The
limiting rotation arises out of viscous dissipation  and r-mode
increases due to gravitational wave emission. Competition between
these two factors determines the stability status, ie, the r-mode
instability window (RMIW).  The critical rotation frequency for a
given stellar model can be derived from following equations
\begin{equation}
\frac{1}{\tau_{gr}}+\frac{1}{\tau_{\nu}}=0,
\end{equation}
where $\tau_{gr}<0$ is the characteristic time scale for energy
loss due to gravitational waves emission, $\tau_{\nu}$ denotes the
damping timescales due to shear, bulk viscosities and other
rubbings. As $\tau_{\nu}$ relates to the viscosities of the matter
inside CS whose differences could result in diverse behaviors.
This could be also used for detecting the interior of CSs.

The bulk viscosity of NS comes from the $\beta$-decay process of
normal nuclear matter. It is commonly considered that the modified
URCA process play a key role (\cite{Sawyer:1979}), but the direct
URCA process(\cite{Haensel:1992}) may become more important when
the temperature of the core in neutron star is high enough.
However, Wang and Lu (\cite{Wang:1984}) pointed that the
non-leptonic weak reaction of quarks had stronger damping effcet
for the radial oscillation of strange quark matter (SQM), which is
to $\sim$ ms order. This means that the bulk viscosity of SQM is
much more larger than normal nuclear matter.
 Sawyer (\cite{Sawyer:1989}), Madsen (\cite{Madsen:1992}) and Goyal et
al.(\cite{Goyal:1994}) did a series of research on the bulk
viscosity coefficient of SQM using bag model. Moreover, Zheng and
his group considered the influence of medium effect on the SQM
bulk viscosity, which restrained the r-mode instability more
effectively(\cite{Zheng:2002};\cite{Yang:2003};\cite{Zheng:2003};\cite{Yu:2004};
\cite{Zheng:2004};\cite{Pan:2005};\cite{Zheng:2005};\cite{Pan:2006};\cite{Zheng:2006}).
Madsen (\cite{Madsen:1998}) just figured that the diversity
between SQM and NS matter could behave in the maximum rotation
frequency. By comparing the shear and bulk viscosity, he found
that the RMIW of strange star(SS) moved to the direction of low
temperature and the new-born SSs may have $\sim$ ms rotation
period, but NSs can't. Meanwhile, Lindblom et
al.(\cite{Lindblom:1998}) and Andersson et
al.(\cite{Andersson:1999}) studied the RMIW of NSs, and recovered
that the viscosity dissipation of young NS couldn't restrain the
r-mode instability and the rotation period would reach 10 $\sim$
20ms in one year. Although Bildsten et al.(\cite{Bildsten:2000})
and Andersson et al. (\cite{Andersson:2000}) concluded that NS
with a solid crust could also explain the conglobation of  Low
mass X-ray Binaries (LMXBs) and the faster rotation periods due to
the larger rubbing in viscosity boundary layer (VBL), the
immediate formation of VBL just after the birth of NS and the
stable existence are difficult to achieve as the heat of viscosity
damping  would destroy the solid configuration of VBL and the
damping would disappear as well. Therefore, it is believed for a
time that if we can discover a pulsar with much shorter period, it
may be SS. Actually, we haven't found pulsars symmetrically on the
left side of the RMIW for SS, which is in contradiction with the
prediction(\cite{Pan:2005}). Moreover,
Jones(\cite{Jones:2001a};\cite{Jones:2001b} )pointed out that the
contribution of hyperons had omitted and then Lindblom et
al.(\cite{Lindblom:2002}) claimed that the non-leptonic reaction
of hyperon matter could also produce larger bulk viscosity and had
stronger restraining effect on r-mode instability. The hybrid star
(HbS) model is also important for studying. Drago et al.
(\cite{Drago:2005} )and Pan et al.(\cite{Pan:2006}) posted that
the non-leptonic reactions of hadron and quark matter in mixed
hadron-quark phase had the same effect
 with the homogeneous neutral phases in despite that it had condensation, superfluidity and
 superconductivity or not

 .Therefore, HbS and
 hyperon star (HpS) are both the best candidates of $\sim$ ms faster rotators according to RMIW.
Recently Kaaret et al. reported their discovery of burst
oscillation at 1122Hz in the X-ray transient XTE J1739-285
(\cite{Kaaret:2007}), which is interpreted as due to the rotation
of the center neutron star and would be the fastest rotating CS
yet found. This puts a much more stringent limit . Let the CS
sequence as the input of equations (1) and (2), we immediately
obtain both the $M-R-\nu_{K}$ and $M-R-\nu_{R}$ relations.
Compared with all observational data, we find that this pulsar can
only obviously be a HbS under Gibbs construction. Although the
result is obtained for specific parameters, this may be a true
signal of quark matter in CSs.

Moreover, thermal emission data of pulsars gives us additional
constraint on CS models. Linking spin evolution to thermal
evolution will present another way to distinguish quark stars from
neutron
stars(\cite{Yu:06};\cite{Zheng:061};\cite{Zheng:062};\cite{Zhou:07};\cite{Kang:07}).

\begin{acknowledgements}
This work is supported by NSFC under Grant Nos.10603002.
\end{acknowledgements}

\end{document}